\documentstyle[twocolumn,psfig]{mn}
\newcommand{\mincir}{\raise -2.truept\hbox{\rlap{\hbox{$\sim$}}\raise5.truept
\hbox{$<$}\ }}
\newcommand{\magcir}{\raise -2.truept\hbox{\rlap{\hbox{$\sim$}}\raise5.truept
\hbox{$>$}\ }}
\newcommand{\minmag}{\raise-2.truept\hbox{\rlap{\hbox{$<$}}\raise 6.truept\hbox
{$>$}\ }}
\newcommand{\be}{\begin{equation}}
\newcommand{\ee}{\end{equation}}
\newcommand{\ba}{\begin{eqnarray}}
\newcommand{\ea}{\end{eqnarray}}
\newcommand{\brr}{\begin{array}}
\newcommand{\err}{\end{array}}
\newcommand{\bc}{\begin{center}}
\newcommand{\ec}{\end{center}}

\title{A possible theoretical explanation of
metallicity gradients in elliptical galaxies}
\author[Martinelli et al.]
{Agostino Martinelli,$^{1}$ Francesca Matteucci,$^{2}$ and Sergio 
Colafrancesco$^{3}$\\
$^1$Instituto Astronomico, Via Lancisi 29  ROMA Italy\\
$^2$Dipartimento di Astronomia, 
Universit\`{a} di Trieste and SISSA Institute, via Beirut 2/4, I--34014 Trieste, Italy\\
$^3$ Osservatorio Astronomico di Roma, Via dell'Osservatorio, I-00040 Monte Porzio, Italy}

\begin{document}

\maketitle

\begin{abstract}

Models of chemical evolution of elliptical galaxies taking into account
different escape velocities at different galactocentric radii are presented.
As a consequence of this,
the chemical evolution
develops differently 
in different galactic regions; in particular, we find that
the galactic wind, powered by supernovae (of type II and I)
starts, under suitable conditions,
in the outer regions and successively develops in the central ones.
The rate of star formation (SFR)
is assumed to stop after the onset of the galactic wind in each region.
The main result found in the present work
is that this mechanism
is able to reproduce metallicity gradients, namely the gradients
in the $Mg_2$ index, in good agreement with
observational data. 
We also find that in order to honor the constant
[Mg/Fe] ratio with galactocentric distance, 
as inferred from metallicity indices, a variable initial mass function
as a function of galactocentric distance is required. 
This is only a suggestion since trends on abundances inferred just from metallicity indices are still uncertain.

\end{abstract}

\section{Introduction}

The usefulness of studying elliptical galaxies
comes both from the importance for cosmology of knowing
their past luminosities and colours, and for testing stellar
evolution models and theories on supernova progenitors.

Many features shown by elliptical galaxies required, in the past
years, theoretical explanations. For example,
in order to reproduce the well known relation between mass and
metallicity Mathews and Baker (1971) and Larson
(1974) proposed the existence of galactic winds driven by
supernova explosions (note that with such a way they explained 
also the apparent
lack of gas in these systems).

Successively, a great deal of effort has been devoted in studying 
detailed chemical
evolution of elliptical galaxies 
(Arimoto and Yoshii, 1986, 1987; Matteucci and Tornamb\`e, 1987;
Brocato E., Matteucci F., Mazzitelli I., Tornamb\`e A., 1990;
Angeletti and Giannone, 1990; Matteucci, 1992; Ferrini
and Poggianti, 1993; Matteucci, 1994; Bressan, Chiosi and Fagotto,
1994; Matteucci
and Gibson, 1995; Gibson, 1997; Gibson and Matteucci 1997).
For example, on the basis of supernova driven galactic wind
models, Matteucci and Tornamb\`e (1987) first analyzed
the evolution of Mg and Fe in these galaxies
and predicted that [Mg/Fe]
should decrease as a function of the galactic mass, at variance with
what has been more recently suggested by population 
synthesis studies (Worthey, Faber and
Gonzalez, 1992) indicating
that [Mg/Fe] increases with galactic mass in the nuclei of giant
ellipticals.
Matteucci (1994) pointed out that, 
in order to explain the origin of this relation one
has to assume either that the efficency of star formation is
an increasing function of mass, or that the initial mass function (IMF) 
should favor
more massive stars in more massive galaxies.
In the models of Larson (1974), Arimoto and Yoshii (1987), and Matteucci and
Tornamb\`e (1987), in fact, the star formation efficiency was either assumed
to be constant or to decrease with galactic mass. This last
hypothesis was based upon the fact that star formation is
mainly determined by cloud-cloud collisions. In this frame-work
the time scale for star formation (namely the time required to
consume all the gas in a galaxy by star formation and 
corresponding to
the inverse of the
star formation efficiency) was assumed to be proportional to
the cloud-cloud collision time or to the dynamical free-fall
time, with the consequence of decreasing with the galactic mass.

Concerning the other explanation relative to the IMF, 
the observed variation of the $M/L_{B}$
ratio in elliptical galaxies as a function of their blue luminosity
(Faber and Jackson, 1976; Michard, 1983; Kormendy and Djorgovski, 1989;
van den Marel, 1991; Bender, Burstein and Faber, 1992) seems to argue in favor
of a possible variation of the IMF with galactic mass, although other
possible explanations cannot be excluded.
Renzini and Ciotti (1993), for example, proposed a variable amount and/or
concentration of 
dark matter as a function of luminous mass;
Zepf and Silk (1996) analyzed 
the hypotesis that star formation includes a phase in which
star formation is restricted to massive stars, with the bias toward
high mass stars increasing with galactic mass.

The evidence for a mass-metallicity relation in ellipticals
comes not only from changes in colors with galactic luminosity
(Faber 1972), but also from the linear relationship existing between
the $Mg_{2}$ index as defined by
Faber S.M., Friel E.D., Burnstein D.,
Gaskell C.M. (1985), 
and the internal velocity dispersion $\sigma$
(Dressler A., Lynden-Bell D., Burnstein D., Davies R.L.,
Faber S.M., Terlevich R.J. and Wegner G. 
1987, Bender, Burstein and Faber, 1993).
It is generally assumed, in fact, that these color 
and metallicity index changes indicate
variations in metallicity.

Ziegler (1997) found that determining the $Mg_{b}-\sigma$ ($Mg_b$ is
another metallicity indicator, always referring to Mg, 
as defined in Faber et al. 1985)
relations at various redshifts is a powerful and robust method to measure
the evolution of elliptical galaxies. In his work he points out
two conclusions: first, the stellar population of elliptical cluster galaxies
evolves passively between $z=0.4$ and today, and second, the epoch of
formation of the stars which form the present day ellipticals
occurred at high redshifts( $Z>2$ and the most luminous objects might
have formed even at $Z>4$).

Another important aspect of the $Mg_{b}-\sigma$ relation is its
small scatter. This is usually interpreted as an indication that the process
of galaxy formation was short (see Renzini 1995)

Metallicity gradients are also well known characteristics of elliptical
galaxies (Carollo Danziger and Buson, 1993; Davies, Sadler and Peletier, 1993)
The best evidence comes from the spectroscopic 
measurements of absorption lines in ellipticals (Faber 1977; Burnstein D.,
Faber S. M.,Gaskell C. M.,
Krumn N, 1984; Faber et al. 1985; Peletier 1989).
Typically, the Mg $b$ and Fe absorption-line
strengths are stronger in high-mass galaxies. In individual
galaxies they generally decrease outward from the center. The colors
behave similarly, being redder in high-mass galaxies and changing from
red to blue outwards. Therefore, both the color and line strength
changes indicate consistently that these gradients are very likely
due to variations in metallicity.
However,
the mechanism by which these metallicity variations could have arisen
is still unclear. The models proposed are based on the processes
developed during galaxy formation. Larson (1974) and Carlberg (1984)
showed that elliptical galaxies which form by dissipative collapse
of a primordial gas gas cloud, should have an increasing metallicity 
with decreasing
radius. The idea is that the stars begin to form everywhere 
during the collapse
and remain in orbit
with a little inward motion, whereas the gas
falls towards the center
and 
is enriched by the evolving stars. In this way, the stars which form at the galaxy center are more metal enriched than those formed in the external regions.
These models produce galaxies with very steep metallicity gradients:
a factor of 10 reduction in metallicity
for a factor of 10 increase in radius.
Carlberg's models include a pressure term associated with the
energy input from supernovae during the collapse. This
decreases the size of abundance gradients in the final galaxies to
a factor of 3 over each decade in radius.

Franx and Illingworth (1990) found that, for their
sample of 17 galaxies, the local (B-R) (U-R) colors are functions
of the local escape velocity $v_{esc}$ for all of the galaxies.
This result was confirmed later by  
Davies Sadler and Peletier (1993) and Carollo and Danziger (1994)
using a more direct indicator of metallicity ($Mg_{2}$) than colors
and a more
appropriate calculation of $v_{esc}$ for each galaxy.
This relation suggests that metallicity gradients
could have arisen because of the
different times of occurence of galactic
winds in different galactic regions. In fact the
galactic wind starts when the energy injected in the
interstellar medium (ISM) by supernovae (of both type I and II)
becomes equal to the binding energy of gas (Larson 1974).
Moreover, it is
reasonable to think that, when the galactic wind starts,
the SFR breaks down.
Hence, in the regions where $v_{esc}$ is low (i.e. where the
local potential is shallow), the galactic wind will develops 
earlier and the gas is less processed than
in the regions where $v_{esc}$ is higher, and the potential
well in the external regions is lower than in the internal ones.

The aim of this paper is to discuss the formation of abundance gradients
in elliptical galaxies.
The model suggested here to explain metallicity gradients
does not 
follow in detail the processes occurred during galaxy formation such as
collapse or merging,
and it starts with
the total galactic mass already present at the beginning.
This is a good assumption for elliptical galaxies since they must have
formed in a reasonably short timescale (Matteucci 1994,
Renzini 1995, Ziegler 1997), as already discussed before.
Recently, Kauffman and Charlot (1997) have discussed the formation of 
ellipticals by merging of spirals in the context of a hierarchical 
clustering scenario. This mechanism would imply a much longer process than
that adopted here and it certainly can not explain the high [Mg/Fe] ratios
found in the nuclei of ellipticals.Moreover, recently Prochaska and 
Wolfe (1997) showed that Kauffman (1996) model is inconsistent
with the damped Lyman-$\alpha$ data.

The results we present here are obtained using
a code of chemical evolution
(Matteucci and Gibson 1995, hereafter MG)
modified in order to take account the different
evolution of each galactic region.

The work is organized in four sections.
Section 2 concerns the new theoretical prescriptions with
respect to MG;
the theoretical results are discussed and compared
with observations in section 3, where, models of a galaxy
with a mass of $10^{11} M_{\odot}$ are discussed for illustrative
purposes.
Finally, some conclusions and remarks are done in section 4.

\section{Theoretical prescriptions}

\subsection{The chemical evolution model}

In the following, we consider elliptical galaxies
to have a spherical symmetry. 
We assume that at the initial time (t=0) these systems consist
only of primordial gas
and the star formation starts
at the same time in all galactic regions.

The galaxy is partitioned in zones having the shape of spherical shell.
The basic equations to calculate the chemical evolution in each zone
are the same as in MG and,
for this reason, they will not be repeated here.
We will only remind that the model belongs to the category of supernova driven
wind models and that follows the evolution in space and time
of the abundances of several chemical
elements (H, D, $^{3}He$, $^{4}He$, C, N, O, 
Ne, Mg, Si, Fe and others).
The stellar lifetimes are taken into account and the rates of supernovae 
of type II and I are computed in detail. In particular, for the rate of type 
Ia supernovae, which are assumed to originate in white dwarfs in binary 
systems we
adopt the original formulation described in Matteucci and Greggio (1986) and
Matteucci and Tornamb\`e (1987).

\subsection{Stellar Birthrate Function}

The stellar birthrate function is generally separated into two
independent functions:
\begin{equation}
B(m,t)=\psi(t)\phi(m)
\end{equation}
\noindent 
Following MG, $\psi(t)$ is given by:
\begin{equation}
\psi(t)=\nu \rho_{gas}(t)
\end{equation}
\noindent 
where $\rho_{gas}(t)$ 
is the volume gas density
in the region of interest at the time
t. The parameter $\nu$ represents  the star formation efficiency and is taken
from Arimoto and Yoshii (1986)
$\nu=8.6(M_{gas}(0)/10^{12}M_{\odot})^{-0.115}(Gyr^{-1})$,
and is assumed to be constant with radius.

The IMF is generally expressed as a power law and here we consider:
\begin{equation}
\phi(m)\propto m^{-x}
\end{equation}
\noindent 
normalized to unity in the mass range
$0.1\leq M/M_{\odot}\leq 100$. We have explored three different cases:
\begin{enumerate}

\item $x=0.95$ for all the regions of galaxy;

\item $x=1.35$ for all the regions of galaxy;

\item $x$ variable and in particular increasing with radius.

\end{enumerate}

\subsection{Stellar Yields}

We adopted the same stellar yields as in MG in order to be able to 
compare our results with theirs.
In particular we assumed:
\par
i) for massive stars and supernovae II
($M> 8M_{\odot}$) the yields of Woosley (1987) are adopted,

\par
ii) for low and intermediate mass stars ($0.8 \le M/M_{\odot} \le8$)
we assume the yields of Renzini and Voli (1981).
\par

iii) For type Ia supernovae the yields of Thielemann, Nomoto and Yokoi
(1993) for their model
W7 are used.
The stellar lifetimes are the same adopted by Matteucci and Padovani (1993),
namely:
\par
for $M> 6.6 M_{\odot}$
\par
\begin{equation}
\tau_m=1.2m^{-1.85} + 0.003 \,\,  Gyr 
\end{equation}
\noindent

and for $M \le 6.6 M_{\odot}$
\par

\begin{equation}
\tau_m= 10^{1.338-\sqrt{1.790-0.2232 \cdot(7.764-logm)}} -9 \,\, Gyr 
\end{equation}
\noindent

\smallskip
For the details about calculation of the type Ia SN rate we address the reader
to MG95. We will remind here only that the fraction of binary systems is a function of the assumed IMF as described in MG95.

\subsection{Galactic Wind}

Galactic winds in ellipticals were originally introduced to reproduce 
the well known mass-metallicity relation in these systems and to explain 
the apparent lack of gas (Larson 1974).
However, besides this, galactic winds are the natural consequence of 
a phase of intense star formation.  
The existence of a galactic wind, in fact,  is strictly related to the
energetics of the interstellar medium compared to the potential 
energy of the gas.
The time at which  a galactic wind occurs is a fundamental parameter
for the chemical evolution of a galaxy region. The
condition which should be satisfied for the gas to
be ejected  from a given region is expressed by the following
condition (Larson 1974):

\begin{equation}
E_{th_{SN}}^{i}(t_{GW})=E_{Bgas}^{i}(t_{GW})
\end{equation}
\noindent 
where the index $i$ indicates the particular region under consideration.

\noindent 
The condition requires that the thermal energy of the gas in a given region,
resulting from supernova explosions in that region, must
exceed the binding energy of the gas.
For both quantities in equation (6) we used
new theoretical prescriptions with respect
to previous works. These prescriptions will be described in 
the next paragraphs.

\subsection{The potential energy of the gas, $E_{Bgas}$}

We define
$E_{Bgas}$ as the energy necessary to carry the gas of a given
galactic region outtside the galaxy.

With reference to a spheric shell region of tickness
$\Delta R_{i}$ and internal and external radii
$R_{i}$ and $R_{i+1}=R_{i}+\Delta R_{i}$ respectively, the energy $E_{Bgas}$
can be written
as follow:

\begin{equation}
E_{Bgas}^{i}=\int_{R_{i}}^{R_{i}+\Delta R_{i}}{dL(r)}
\end{equation}
\noindent
where $dL(r)$ is the energy required to carry out a quantity
$dm=4\pi r^{2}f_{g}(r,t)dr$ of gas. Here $f_{g}(r,t)$ is the density
function of gas, depending on the distance r from the galactic
center and on the galactic age and t. Therefore, we have for $dL(r)$:

\begin{equation}
dL(r)=\int_r^{\infty}{dr'f(r')}
\end{equation}
\noindent
where $f(r')$ is the force between $dm$ and the total mass
(dark and luminous) within r' which reads as:

\begin{equation}
f(r')=\frac{GM(r')dm}{r'^2}
\end{equation}

\noindent
The range of integration in (8) can be
reasonably limited to an upper effective radius, 
since we only require that the
gas leaves the galaxy. For this reason we replace $\infty$
with $(1+a)R_e$ , where $R_e$ is the effective galaxy radius and
$a\geq0$.  
As a consequence of this,
$dL(r)=\int_r^{(1+a)R_e}{dr'f(r')}$.
In practice, we have analyzed two cases:
i) $a=0.5$ which corresponds to the case
in which the gas is carried out at $1.5R_{e}$, and
ii) $a=\infty$ which corresponds to the case in which the gas is 
carried out to infinity.
The influence of a finite value of $a$ will be discussed
in section 3.

The total mass at radius r', $M(r')$,  is given by:

\begin{equation}
M(r')=M_{Dark}(r')+M_{Lum}(r')
\end{equation}

\noindent
where the luminous mass can be written as the
sum of the gaseous and stellar components:

\begin{equation}
M_{Lum}(r')=M_{gas}(r')+M_{*}(r')
\end{equation}

\noindent
Assuming the same distribution for the two components, we have:

\begin{equation}
M_{gas}(r')=\rho_{g}F_{l}(r');
\end{equation}

\begin{equation}
M_{*}(r')=\rho_{*}F_{l}(r');
\end{equation}

\begin{equation}
F_{l}(r')=4\pi \int_0^{r'}{dsf_{l}(s)s^2};
\end{equation}

\noindent
The quantities appearing in equations  (12), (13) and (14)
can be determined on the basis of
prescriptions available in the literature. Namely, we have for
$F_{l}(r)$ (Jaffe 1983)

\begin{equation}
F_{l}(r)\propto\frac{r/r_.}{1+r/r_.}
\end{equation}

\noindent
with $r_.=R_e/0.763$, and
for the dark matter distribution we have (Bertin Saglia and Stiavelli, 1992)

\begin{equation}
M_{Dark}(r)=\frac{2M_{Dark}}{\pi}[arctg(r/r_0)-\frac{r/r_0}{1+(r/r_0)^2}]
\end{equation}

\noindent
with $r_0=0.45R_{Dark}$. In the following, three different
cases for the distribution of dark matter
will be analyzed:

\begin{enumerate}

\item $M_{Dark}=0$

\item $M_{Dark}=10M_{Lum}$ and $R_{Dark}=10R_{e}$

\item $M_{Dark}=10M_{Lum}$ and $R_{Dark}=R_{e}$

\end{enumerate}
These cases have been chosen since they represent extreme cases going from no
dark matter (except the stellar remnants) to an amount of dark matter
ten times the luminous one either distributed in a diffuse halo
or tracing the luminous material. All the other possibilities will therefore be
intermediate between these cases.

In the case $a=\infty$, $R_i=0$,
$R_{i+1}=R_e$ (i.e. one-zone model), considering
luminous matter uniformly distributed and in absence of
dark matter, eq. (7) gives:

\begin{equation}
E_{Bgas}=\frac{3}{5}\frac{G}{R_e}M_{gas}(2M_*+M_{gas})
\end{equation}
\vskip 0.5truecm

\subsection{The thermal energy of the gas due to supernovae, $E_{th_{SN}}$}

The total thermal energy in the gas at the time $t$ is:

\begin{equation}
E_{th_{SN}}(t)=\int_0^t{\epsilon_{th_{SN}}(t-t')R_{SN}(t')dt'}
\end{equation}

\noindent 
where $R_{SN}(t)$ is the SN rate and $\epsilon_{th_{SN}}(t_{SN})$ is
the thermal energy content inside a supernova remnant of
both types, with $t_{SN}=t-t^{'}$. $\epsilon_{th_{SN}}(t_{SN})$
evolves, according to Cox (1972), in the following
way:

\begin{equation}
\epsilon_{th_{SN}}(t_{SN})=7.2 10^{50}\epsilon_0  ergs
\end{equation}

\noindent
for $0\leq t_{SN}<t_c$, and

\begin{equation}
\epsilon_{th_{SN}}(t_{SN})=2.2 10^{50}\epsilon_0(t_{SN}/t_c)^{-0.62}ergs
\end{equation}

\noindent
for $t_{SN}\geq t_c$

where $\epsilon_0$ is the initial blast wave energy in units
of $10^{51}$ ergs, $t_{SN}$ is the time elapsed since the supernova
explosion and $t_c$ is the cooling time of a supernova remnant,
namely

\begin{equation}
t_c=5.7 10^{4}\epsilon_0^{4/17}n_0^{-9/17}yr
\end{equation}

\noindent
and $n_0(t)=\rho(t)/<m>$ is the average number density of the ISM.
Gibson (1997) adopted a more sophisticated 
treatment for the calculation of $E_{th_{SN}}$, based on calculations by 
Cioffi McKee and Bertschinger (1988). However, given the uncertainties present 
in this kind of 
calculations we have decided to adopt the old formulation, common to 
most of the previous papers on the chemical evolution of ellipticals.
We did not consider the energy injected into the ISM by stellar winds,
since for massive galaxies is negligible, as shown by Gibson (1994).

\section{Results and conclusions}

\subsection{Model parameters}

We have calculated several chemical evolution models by varying the amount and
distribution of dark matter as well as the parameter $a$ 
and we have also considered either the case of a constant
IMF
in all the regions of the galaxy or
the case of an IMF varying with the galactocentric
distance.
In Table 1 we show the choice of the
parameters for all the models referring to the case with constant IMF.
In particular, in this table we indicate the amount of dark matter 
(column two) and the distribution of dark matter (column three). 
Finally, in column four we show the adopted value of the parameter $a$.
We have used two different values for the IMF slope
in the models with an IMF costant in radius: $x=0.95$ and
$x=1.35$ (Salpeter 1955).

In the framework of the models with a variable IMF
we considered only one case in which all the
parameters are the same as in the B2 model, except for the IMF slope.
We will refer to this last case as the model IV, where the IMF
slope varies from $x=0.95$ in the internal
regions to $x=1.21$ in the external
ones. This choice is made in order to
have the $[Mg/Fe]$ constant in radius
as it will be shown. Observational results seem, in fact,
to indicate that $[Mg/Fe]$ as a function of the galactocentric
distance is roughly constant, at variance with what happens for the
central $[Mg/Fe]$ as a function of galactic mass 
(Worthey et al. 1992;
Carollo et al. 1993; Davies et al.1993). 
However, there is a warning here, namely the trends on real abundances 
inferred from metallicity indices may be deceiving as recently
suggested by Tantalo Bressan and Chiosi,
(1997) and Matteucci and Ponzone (1997).
In all the cases the values of $t_{GW}$, the iron
abundances and the $[Mg/Fe]$ in the dominant stellar population will
be calculated for a galaxy of $10^{11} M_{\odot}$ of luminous mass and
$R_e=3Kpc$.

\subsection{Galactic winds}

The results concerning galactic winds are summarized in Fig. 1
and Tables 2-14.
Fig. 1 shows the behaviour of $E_{Bgas}$ (solid line) and
$E_{th_{SN}}$ (dashed line) for the 6 one-zone models.
The occurence of galactic winds is affected by the presence
of dark matter depending on the value of $a$. 
The value of $a$ sets the distance at which the gas is ejected. Small values
of $a$ imply the presence of gas near the galaxy whereas large values of $a$
imply that the gas is carried outside the gravitational field of the
galaxy. This formulation allows us to study different cases unlike the
model of Matteucci (1992) which corresponded only to the case of a small
$a$.
Although a dynamical model would be necessary to assess if the gas mass 
is definitely lost from the galaxy or if it remains around it, 
different values of the parameter $a$ may influence the chemical 
enrichment of the ICM, in the
sense that large values of $a$ will predict a higher contribution 
from the galaxies to the gas of the ICM.
To be more precise the influence
of dark matter is very strong for all values of $a$ when it is distributed
as the luminous one (models C1 and C2). However, when dark matter is
distributed in a diffuse halo, its effect is relevant only for high
values of $a$ (B1): in fact, in this case, we are considering the potential
energy of the gas as the energy to carry the gas out the halo. On the other
hand, a diffuse halo of dark matter does not affect the occurence of
galactic wind if we use small values of $a$ (B2). This last result is in
agreement with Matteucci (1992): she defined the potential energy
as the energy to carry the gas out the galaxy, namely at the distance
$R_e$ ($a=0$).



Tables 2-14 give the values of $t_{GW}$ for the models
in the case of 10 zones with IMF slope equal to $0.95$ (tab. 2-7),
$1.35$ (tab. 8-13) and variable (tab. 14)
(the zone tickness are obtained
directly from data observation by Davies et al.,1993).

>From these tables we note that galactic winds are present only
in a small number of outer regions for models C.
Namely, the wind occurs only for $r > 1.5 Kpc$
and $r> 1.0 Kpc$ for models C1 and C2 respectively.
Consequently, models C predict much smaller gas masses to be ejected
from the galaxy and masses to form the hot 
coronae smaller than the those  estimated
from the X-ray emission (Forman Jones and Tucker, 1985; Canizares 
Fabbiano and Trinchieri, 1987).
Such galaxies should also show active star formation at intermediate redshifts
($z<1$), which seems not to be the case for normal ellipticals
(Matteucci 1994; Renzini 1995; Ziegler 1997).
Therefore, we are tempted to reject the hypothesis of dark matter
distributed like the luminous one (as already pointed out by Matteucci 1992).

With regards to models A1, A2, B2, we note they are very similar to those
in the case of one-zone. For these models, galactic winds occur
in the internal regions after several Gyr since galaxy formation
(note that in the most internal region, with $r<30 pc$, the wind never occurs),
while in the most external regions they occur before 1.0 Gyr from the epoch of
galaxy formation.

This time spread of the galactic wind occurence with the
galactocentric distance, has very important consequences
on the chemical evolution in clusters of galaxies.
Previous one zone models, in fact, including massive, diffuse,
dark matter component (B2 models), undergo only a global "impulsive"
ejection of their ISMs at $t_{GW}$
(the same result was pointed out by MG),
also for different prescriptions of
the IMF and SFR). Therefore, these one-zone models do not predict
any chemical evolution in the ICM after nearly $1 Gyr$ from galaxy
formation, when
galactic winds have already occurred in all the galaxies.
On the other hand, the multi-zone models predict chemical evolution in the
ICM also at the present time because, even if there is only one global 
"impulsive" ejection of gas in all the regions of the galaxy,
the most internal regions eject the processed gas very late.
Another consequence, which we will see in the next paragraphs is the predicted
behaviour of [Mg/Fe] in stars which tend
to increase with increasing radius, unless a variable IMF is assumed.
This could be a potential problem since the galactic wind occurring 
late in the most
internal regions of ellipticals might mean active star formation at
small red-shifts, 
which is not observed in these galaxies. However, as it is shown in Fig. 2 
the SN II rates (and therefore the SFR) predicted by our best model 
IV starts to be negligible  
after $t> 3$ Gyr.
This is due to the fact that the gas is consumed by star formation much before
the galactic wind occurs.
We do not show the SN rates for our model B2(x=0.95) since they are very
similar to the results of model IV.


\subsection{Metallicity in stars}

According to the model of chemical evolution used, each
region of elliptical galaxies has a composite
stellar population containing stars of different ages and metallicities.
This is, of course, true also in reality since the process of star
formation is not instantaneous. In order to compare theoretical results
with the metallicity abundances indicated by the observed indices
we should first calculate the average iron and magnesium
abundances in the composite stellar populations.
The mass-averaged metallicity of a composite stellar population
is defined following Pagel and Patchett (1975) as:

\begin{equation}
<Z>_{M}=\frac{1}{S_{1}} \int_{0}^{S_{1}}{Z(S)dS}
\end{equation}
\noindent
where the subscript 1 indicates a specific time $t_{1}$, which in
our case is the present time, assumed to be 15 Gyr
and $S$ is the total
mass of stars ever born. According to the definition above
we can calculate the average iron and magnesium abundances, since
the model provides the functions $X_{Fe}(S)$ and $X_{Mg}(S)$. However,
it should be noted that, although the mass-averaged
metallicity represents the real average metallicity of the composite stellar
population, what should be compared with the metallicities deduced from
observations is the luminosity-averaged metallicity.
In fact, the metallicity indices, measured from integrated spectra,
represent the metallicity of the stellar population which
predominates in the visual light.
The transformation from metallicity indices to real abundances
should then be performed through either theoretical or
empirical calibrations relating metallicity indices to real abundances.
Galactic chemical evolution models
(Arimoto and Yoshii, 1987; Matteucci and Tornamb\`e, 1987) have shown that
the two averaged metallicities are different, since metal poor
giants tend to predominate in the visual luminosity so that, in general, the
luminosity-averaged metallicity is smaller than the mass-averaged one.
However, the difference is negligible for galaxies with mass $M \ge 
10^{11} M_{\odot}$ which is the case we discuss here, and
we are interested mainly in studying metallicity
gradients which are clearly
not affected by the absolute value of the metallicity.

The results concerning the abundance of Fe
in the dominant stellar population and the indices are summarized in
the 3-d and 4-st columns of tab. 2-14. In these same tables we report
the logarithm of (${R \over R_e}$)
(column one) and the time for the
onset of galactic winds expressed in Gyr (column two).
In all the tables but table 14 column 5 shows
the [${Mg \over Fe}$] ratio in the dominant stellar population. 
In table 14 instead we show in column five the adopted
values for the IMF slope as functions of the galactocentric distance.
In this model [${Mg \over Fe}$]
is constant
as a function of the galactocentric distance 
($[Mg/Fe]=0.15$), since we selected the 
IMF slope in order to achieve this result.
The [Fe/H] of the dominant stellar population (column 3)
is 
transformed into
$Mg_2$ (column 4)
by using the calibration of Worthey (1994) corresponding to an age of the 
dominant stellar population of 17 Gyr and [Mg/Fe]=0.

In models A2 and B2 with $x=1.35$
we obtain for $[\frac{Fe}{H}]$ values of 0.55
and 0.025 respectively in the central and external regions. Correspondingly
we obtain a gradient of  $\frac{\triangle{[Fe/H]}}{\triangle{lgr}}$= 0.27.
We tried to use also other calibrations available in the literature
(Buzzoni Gariboldi and Mantegazza, 1992; Mould 1978; Barbuy, 1994) and we found
differences either in the absolute values of $Mg_{2}$ 
up to $\sim 0.1$ dex.
\par
In fig. 3 the predicted values of $Mg_2$ versus radial distance
from galactic center are shown
together with data from Davies et al. (1993).
The data refer to galaxies of the same size as our typical galaxy.
In the figure the solid line represents the predictions of model IV,
whereas dotted and dashed lines
represent the predictions of model  B2 with 
$x=0.95$ and $x=1.35$, respectively.
Theoretical values are sistematically larger than
observational data. This fact can be due 
to the uncertainties related to the metallicity calibration and 
to the stellar yields.
On the basis of
the previous remark it is interesting to note
that a better matching between theoretical and experimental
data can be simply obtained by shifting the theoretical curves
by a suitable amount.
The shifts we applied are -0.079 dex for model IV, -0.084 for
model B2 (0.95) and -0.05 for model B2(1.35)
These shifts are well inside the differences we 
obtained by changing
calibrations
and/or by using models of different ages
as it is shown in fig. 4 where the predictions of our best model (model IV)
are transformed according to the calibratrtions of Worthey (1994) for
different ages of the dominant stellar populations and [Mg/Fe]=0.
In any case,
we do not think that is very important to fit the absolute
values of $Mg_2$ giving the still existing uncertainties, but that 
is more meaningful to fit the gradient itself and our models produce
a very good fit.
In particular,
the model with $x=1.35$ shows a metallicity gradient
in very good agreement with data , but the value of $[Mg/Fe]$ 
in the central region is
$-0.3$, namely too low. In fact, the values suggested by population 
synthesis studies (Worthey et al. 1992; Weiss, Peletier and Matteucci, 1995)
indicate positive values for this ratio.
This result is mostly due to the late occurrence of galactic winds 
in the center of the galaxy which 
allow the supernovae of type Ia to restore the bulk of iron 
before star formation stops. 
However, as discussed by Gibson (1997), yields different than 
those adopted here may increase the value of [Mg/Fe] but only on 
short timescales, namely
when the contribution of massive stars is still dominant. In fact
at late times, when the SNe Ia are appearing, the [Mg/Fe] ratio decreases 
because of the newly injected iron. 
In addition, this model predicts that [Mg/Fe] should increase 
with increasing radius, clearly at variance with what inferred 
from metallicity indices.
On
the other hand, model B2 with $x=0.95$ shows a metallicity gradient
generally flatter than the observational data  but 
a better ratio [Mg/Fe]
($+0.15$ in the central region).
However, the best model seems to be the model IV with a variable
IMF which shows a good
metallicity gradient and a correct value of $[Mg/Fe]$
in the center and in the other galactic regions.
This best model predicts also SN rates
(expressed in units of SNu, namely SNe 
$century^{-1}$ $10^{-10} L_{B_{\odot}}$) 
and blue luminosity at the present time in good agreement with observations
(Turatto, Cappellaro and Benetti, 1994),
namely $(R_{SNIa})_{centre}=1.9$ SNu, $(R_{SNII})_{centre}= 0$ SNu 
and $(R_{SNIa})_{R_e}=0.17$ SNu, 
$(R_{SNII})_{R_e}= 0$ SNu. 
The predicted integrated blue luminosity  
for the whole galaxy, calculated as in Matteucci (1994),  
is $L_{\odot}=7 \cdot 10^{9}L_{B_{\odot}}$.

\section{Conclusions and Discussion}

In this paper we have explored the possibility of explaining metallicity 
gradients in giant ellipticals by means of models of chemical evolution
including galactic winds. 
We have analyzed the case of an elliptical galaxy with initial luminous mass 
$M_{lum}=10^{11} M_{\odot}$, $R_{e}=3$ Kpc and with a dark matter halo
ten times more massive than the luminous mass.

We found that, in order to reproduce most of the observational constraints
relative to elliptical galaxies:
\par
- the dark matter should not be distributed like the luminous matter but
it should be more diffuse.
This result reinforces a
previous analysis made by Matteucci (1992).
\par
-Abundance gradients in the stellar populations of ellipticals can be 
obtained by assuming that galactic winds develop first in the outer 
and later in the inner galactic regions, in agreement with the 
observational evidence of a correlation between the gas escape 
velocity and  metallicity
(Franx and Illingworth, 1990;
Davies, Sadler and Peletier, 1993; Carollo and Danziger, 1994).  
The physical justification for this
resides in the fact that the force required to carry out
mass from the more external regions is lower than that required to carry out
mass in the more internal ones.
This mechanism for the formation of gradients is different from the mechanism 
often invoked (Larson, 1974; Carlbergh 1984; Greggio 1997). 
In this framework, an abundance gradient should 
arise because 
the stars form everywhere in a collapsing clouds and then remain in orbit 
with a little inward motion whereas the gas sinks further in because there 
is dissipation.
This gas contains the metals ejected  by evolving stars so that 
an abundance gradient develops in the gas.
As stars continue to form their composition reflect the gaseous 
abundance gradient.

Our model does not exclude a mechanism of this type which can be present
together with the galactic winds occurring at different epochs in 
different regions. 
What we want to show here is that the differential occurrence of 
galactic winds alone can explain the observed abundance gradients.
On the other hand, a mechanism of inside-out
galaxy formation, as that required
to reproduce abundance gradients in the disk of our Galaxy
(Matteucci and Fran\c cois, 1989: Chiappini, Matteucci and Gratton 
1997) , does not seem to be 
suitable for elliptical galaxies. In fact, in order to reproduce the 
abundance gradient in the disk of our Galaxy one needs a large delay 
between the formation, due to infall of gas, 
of the more internal and the more external parts 
of the disk, of the order of several
billion years (Chiappini et al., 1997). 
This would imply that most ellipticals should be still forming now
with active star formation
like in the disks of spirals, at variance with observations.
The advantage of the model proposed here over other models such as that of
Larson (1974) is that it explains the observed correlation between 
colors/metallicity and escape velocity 

\par
A final remark concerns the star formation efficiency which we have
assumed to be constant with galactocentric distance.
We have also tried to vary this efficiency with galactocentric distance 
in the sense of having a higher efficiency in the central regions.
We found that for moderate variations of the efficiency of star formation 
the abundance gradients steepen together with gradient in [Mg/Fe].
On the other hand, if the variation of $\nu$ is too strong the trend for
the galactic wind is reversed, occurring later in the outermost regions with
the consequence of cancelling the abundance gradients.

\par
- However, a potential problem with the mechanism
for the formation of gradients suggested here
is that it would predict
an increasing [Mg/Fe] as a function of the galactocentric distance,
at variance with observations, unless a 
variable IMF is assumed at the same time.
\par
- In order to obtain an almost constant [Mg/Fe] as a function of the 
galactocentric distance, as suggested by observations (Worthey
et al., 1992; Carollo et al., 1993; 
Carollo and Danziger 1994), 
one has also to
assume that the IMF varies with radius. 
In particular, $x$ should vary 
from 0.95 at the galaxy center up to 1.21 in the outermost regions.
This would imply a slightly higher percentage of massive stars in the 
innermost regions thus furtherly favoring the abundance gradients.
However, the observed trend of [Mg/Fe] is inferred from the metallicity
indices which are also affected by other parameters such as age.
Therefore, 
it is very important to have reliable [Mg/Fe] ratios as a function of
the galactocentric distances in ellipticals in order to assess this point.

-The required IMF, in order to obtain positive values of [Mg/Fe], 
especially in the galaxy center, as suggested by the observations (Worthey
et al., 1992; Weiss et al., 1995),
must have a slope $x=0.95$ 
in agreement with previous studies (MG; Gibson and Matteucci, 1997), 
although a variation in the stellar yields could achieve the same result 
(Gibson 1997) but only if the timescale for the galactic wind is 
short enough (Weiss, Peletier and Matteucci, 1995).
\par

\par

-Another potential problem could be the fact that in our best model
(model IV) the galactic wind does not develop until 12 Gyr in the most
internal regions ($< 150$ pc).
However, it should be noted that due to the fast consumption of gas
star formation occurs at such a  low level at 12 Gyr that would be 
hardly detectable.
\par

-Finally, we want to discuss the consequences of our model on the chemical
evolution of the intracluster medium (ICM) although more 
details will be found
in a future paper (Colafrancesco S.,  Matteucci F., Martinelli A.,
and Vittorio N., in preparation).
The late occurrence of galactic winds in the more internal regions 
of ellipticals, if true,
will have as a consequence a continuous chemical evolution of the ICM,
at variance with the predictions by models with only early winds 
(MG; Gibson and Matteucci, 1997) which predict no evolution of 
the ICM starting from very
high redshift (Z$ \le 4-5$). This prediction can be checked in the future
with
observations of abundances in high red-shift clusters; preliminary 
results from ASCA (Mushotsky and Lowenstein, 1997) seem to indicate 
no evolution in
the abundance of Fe for $Z>0.3$.

\newpage


\begin{table}[hbt]
\begin{center}
\begin{tabular}{||l|l|l|l|l||}
\hline
MODEL & $\frac{M_D}{M_L}$ & $\frac{R_D}{R_L}$ & $a$  \\
\hline
A1 & $0$  & $/$ & $\infty$ \\ 
\hline
A2 & $0$  & $/$ & $0.5$ \\ 
\hline
B1 & $10$  & $10$ & $\infty$ \\ 
\hline
B2 & $10$  & $10$ & $0.5$ \\ 
\hline
C1 & $10$  & $1$ & $\infty$ \\ 
\hline
C2 & $10$  & $1$ & $0.5$ \\ 
\hline
\end{tabular}
\caption{Models considered. The parameter a, in 3-d column,
fixes the distance at which the gas is carried out the galaxy, namely
$(1+a)R_e$}
\end{center}
\end{table}


\begin{table}[hbt]
\begin{center}
\begin{tabular}{||l|l|l|l|l||}
\hline
$Log(\frac{R}{R_e})$ & $t_{GW}$ & $[\frac{Fe}{H}]$ & $Mg_2$ & $[\frac{Mg}{Fe}]$  \\
\hline
-2.233 & $/$  & $0.68$ & $0.400$ & $0.15$ \\ 
\hline
-1.929 & $/$  & $0.68$ & $0.400$ & $0.15$ \\ 
\hline
-1.385 & $13.5$  & $0.68$ & $0.400$ & $0.15$ \\ 
\hline
-0.963 & $5.0$  & $0.65$ & $0.394$ & $0.17$ \\ 
\hline
-0.824 & $3.6$  & $0.63$ & $0.391$ & $0.18$ \\
\hline
-0.602 & $2.5$  & $0.59$ & $0.385$ & $0.20$ \\
\hline
-0.424 & $1.5$  & $0.55$ & $0.378$ & $0.22$ \\
\hline
-0.279 & $1.1$  & $0.52$ & $0.373$ & $0.24$ \\
\hline
-0.137 & $0.82$  & $0.48$ & $0.366$ & $0.25$ \\
\hline
0.000 & $0.62$  & $0.43$ & $0.357$ & $0.27$ \\
\hline
\end{tabular}
\caption{Model A1 with the IMF slope ($x$)
constant in radius and equal to $0.95$} 
\end{center}
\end{table}


\begin{table}[hbt]
\begin{center}
\begin{tabular}{||l|l|l|l|l||}
\hline
$Log(\frac{R}{R_e})$ & $t_{GW}$ & $[\frac{Fe}{H}]$ & $Mg_2$ & $[\frac{Mg}{Fe}]$  \\
\hline
-2.233 & $/$  & $0.68$ & $0.400$ & $0.15$ \\ 
\hline
-1.929 & $/$  & $0.68$ & $0.400$ & $0.15$ \\ 
\hline
-1.385 & $11.8$  & $0.67$ & $0.400$ & $0.15$ \\ 
\hline
-0.963 & $4.3$  & $0.64$ & $0.393$ & $0.17$ \\ 
\hline
-0.824 & $3.1$  & $0.62$ & $0.389$ & $0.18$ \\
\hline
-0.602 & $1.7$  & $0.56$ & $0.380$ & $0.22$ \\
\hline
-0.424 & $1.1$  & $0.53$ & $0.374$ & $0.24$ \\
\hline
-0.279 & $0.86$  & $0.49$ & $0.367$ & $0.25$ \\
\hline
-0.137 & $0.58$  & $0.42$ & $0.355$ & $0.27$ \\
\hline
0.000 & $0.40$  & $0.31$ & $0.336$ & $0.29$ \\
\hline
\end{tabular}
\caption{Model A2 with $x=0.95$} 
\end{center}
\end{table}


\begin{table}[hbt]
\begin{center}
\begin{tabular}{||l|l|l|l|l||}
\hline
$Log(\frac{R}{R_e})$ & $t_{GW}$ & $[\frac{Fe}{H}]$ & $Mg_2$ & $[\frac{Mg}{Fe}]$  \\
\hline
-2.233 & $/$  & $0.68$ & $0.400$ & $0.15$ \\ 
\hline
-1.929 & $/$  & $0.68$ & $0.400$ & $0.15$ \\ 
\hline
-1.385 & $/$  & $0.68$ & $0.400$ & $0.15$ \\ 
\hline
-0.963 & $9.4$  & $0.67$ & $0.398$ & $0.15$ \\ 
\hline
-0.824 & $7.0$  & $0.66$ & $0.397$ & $0.16$ \\
\hline
-0.602 & $4.5$  & $0.64$ & $0.393$ & $0.17$ \\
\hline
-0.424 & $3.0$  & $0.62$ & $0.389$ & $0.18$ \\
\hline
-0.279 & $2.4$  & $0.59$ & $0.385$ & $0.20$ \\
\hline
-0.137 & $1.7$  & $0.56$ & $0.380$ & $0.22$ \\
\hline
0.000 & $1.3$  & $0.54$ & $0.376$ & $0.23$ \\
\hline
\end{tabular}
\caption{Model B1 with $x=0.95$}
\end{center}
\end{table}


\begin{table}[hbt]
\begin{center}
\begin{tabular}{||l|l|l|l|l||}
\hline
$Log(\frac{R}{R_e})$ & $t_{GW}$ & $[\frac{Fe}{H}]$ & $Mg_2$ & $[\frac{Mg}{Fe}]$  \\
\hline
-2.233 & $/$     & $0.68$ & $0.400$ & $0.15$ \\ 
\hline
-1.929 & $/$     & $0.68$ & $0.400$ & $0.15$ \\ 
\hline
-1.385 & $11.9$  & $0.68$ & $0.399$ & $0.15$ \\ 
\hline
-0.963 & $4.1$   & $0.64$ & $0.393$ & $0.17$ \\ 
\hline
-0.824 & $3.0$   & $0.62$ & $0.389$ & $0.18$ \\
\hline
-0.602 & $1.8$   & $0.57$ & $0.381$ & $0.21$ \\
\hline
-0.424 & $1.2$   & $0.53$ & $0.375$ & $0.23$ \\
\hline
-0.279 & $0.87$  & $0.50$ & $0.368$ & $0.25$ \\
\hline
-0.137 & $0.59$  & $0.43$ & $0.358$ & $0.27$ \\
\hline
0.000  & $0.35$  & $0.31$ & $0.336$ & $0.29$ \\
\hline
\end{tabular}
\caption{Model B2 with $x=0.95$} 
\end{center}
\end{table}


\begin{table}[hbt]
\begin{center}
\begin{tabular}{||l|l|l|l|l||}
\hline
$Log(\frac{R}{R_e})$ & $t_{GW}$ & $[\frac{Fe}{H}]$ & $Mg_2$ & $[\frac{Mg}{Fe}]$  \\
\hline
-2.233 & $/$     & $0.68$ & $0.400$ & $0.15$ \\ 
\hline
-1.929 & $/$     & $0.68$ & $0.400$ & $0.15$ \\ 
\hline
-1.385 & $/$     & $0.68$ & $0.400$ & $0.15$ \\ 
\hline
-0.963 & $/$     & $0.68$ & $0.400$ & $0.15$ \\ 
\hline
-0.824 & $/$     & $0.68$ & $0.400$ & $0.15$ \\ 
\hline
-0.602 & $/$     & $0.68$ & $0.400$ & $0.15$ \\ 
\hline
-0.424 & $/$     & $0.68$ & $0.400$ & $0.15$ \\ 
\hline
-0.279 & $9.9$  & $0.67$ & $0.398$ & $0.15$ \\
\hline
-0.137 & $5.9$  & $0.65$ & $0.396$ & $0.16$ \\
\hline
0.000  & $3.8$  & $0.63$ & $0.392$ & $0.18$ \\
\hline
\end{tabular}
\caption{Model C1 with $x=0.95$} 
\end{center}
\end{table}


\begin{table}[hbt]
\begin{center}
\begin{tabular}{||l|l|l|l|l||}
\hline
$Log(\frac{R}{R_e})$ & $t_{GW}$ & $[\frac{Fe}{H}]$ & $Mg_2$ & $[\frac{Mg}{Fe}]$  \\
\hline
-2.233 & $/$     & $0.68$ & $0.400$ & $0.15$ \\ 
\hline
-1.929 & $/$     & $0.68$ & $0.400$ & $0.15$ \\ 
\hline
-1.385 & $/$     & $0.68$ & $0.400$ & $0.15$ \\ 
\hline
-0.963 & $/$     & $0.68$ & $0.400$ & $0.15$ \\ 
\hline
-0.824 & $/$     & $0.68$ & $0.400$ & $0.15$ \\ 
\hline
-0.602 & $/$     & $0.68$ & $0.400$ & $0.15$ \\ 
\hline
-0.424 & $9.5$   & $0.67$ & $0.398$ & $0.15$ \\ 
\hline
-0.279 & $5.0$  & $0.65$ & $0.394$ & $0.17$ \\
\hline
-0.137 & $2.7$  & $0.61$ & $0.388$ & $0.19$ \\
\hline
0.000  & $1.4$  & $0.54$ & $0.376$ & $0.23$ \\
\hline
\end{tabular}
\caption{Model C2 with $x=0.95$} 
\end{center}
\end{table}


\begin{table}[hbt]
\begin{center}
\begin{tabular}{||l|l|l|l|l||}
\hline
$Log(\frac{R}{R_e})$ & $t_{GW}$ & $[\frac{Fe}{H}]$ & $Mg_2$ & $[\frac{Mg}{Fe}]$  \\
\hline
-2.233 & $/$    & $0.55$ & $0.378$ & $-0.32$ \\ 
\hline
-1.929 & $/$    & $0.55$ & $0.378$ & $-0.32$ \\ 
\hline
-1.385 & $/$    & $0.55$ & $0.378$ & $-0.32$ \\ 
\hline
-0.963 & $5.3$  & $0.50$ & $0.369$ & $-0.29$ \\ 
\hline
-0.824 & $3.6$  & $0.46$ & $0.362$ & $-0.27$ \\
\hline
-0.602 & $2.0$  & $0.35$ & $0.343$ & $-0.19$ \\
\hline
-0.424 & $1.26$  & $0.26$ & $0.329$ & $-0.13$ \\
\hline
-0.279 & $0.90$  & $0.20$ & $0.318$ & $-0.084$ \\
\hline
-0.137 & $0.72$ & $0.14$ & $0.308$ & $-0.044$ \\
\hline
0.000 & $0.56$  & $0.071$ & $0.296$ & $0.0030$ \\
\hline
\end{tabular}
\caption{Model A1 with $x=1.35$}
\end{center}
\end{table}


\begin{table}[hbt]
\begin{center}
\begin{tabular}{||l|l|l|l|l||}
\hline
$Log(\frac{R}{R_e})$ & $t_{GW}$ & $[\frac{Fe}{H}]$ & $Mg_2$ & $[\frac{Mg}{Fe}]$  \\
\hline
-2.233 & $/$    & $0.55$ & $0.378$ & $-0.32$ \\ 
\hline
-1.929 & $/$    & $0.55$ & $0.378$ & $-0.32$ \\ 
\hline
-1.385 & $14.3$ & $0.55$ & $0.378$ & $-0.32$ \\ 
\hline
-0.963 & $4.3$  & $0.48$ & $0.366$ & $-0.28$ \\ 
\hline
-0.824 & $2.9$  & $0.43$ & $0.357$ & $-0.25$ \\
\hline
-0.602 & $1.6$  & $0.31$ & $0.336$ & $-0.16$ \\
\hline
-0.424 & $0.96$  & $0.21$ & $0.320$ & $-0.094$ \\
\hline
-0.279 & $0.72$  & $0.14$ & $0.308$ & $-0.044$ \\
\hline
-0.137 & $0.56$ & $0.071$ & $0.296$ & $0.0030$ \\
\hline
0.000 & $0.43$  & $0.025$ & $0.288$ & $0.030$ \\
\hline
\end{tabular}
\caption{Model A2 with $x=1.35$}
\end{center}
\end{table}


\begin{table}[hbt]
\begin{center}
\begin{tabular}{||l|l|l|l|l||}
\hline
$Log(\frac{R}{R_e})$ & $t_{GW}$ & $[\frac{Fe}{H}]$ & $Mg_2$ & $[\frac{Mg}{Fe}]$  \\
\hline
-2.233 & $/$    & $0.55$ & $0.378$ & $-0.32$ \\ 
\hline
-1.929 & $/$    & $0.55$ & $0.378$ & $-0.32$ \\ 
\hline
-1.385 & $/$    & $0.55$ & $0.378$ & $-0.32$ \\ 
\hline
-0.963 & $11.8$  & $0.54$ & $0.377$ & $-0.32$ \\ 
\hline
-0.824 & $7.4$  & $0.52$ & $0.373$ & $-0.31$ \\
\hline
-0.602 & $4.4$  & $0.49$ & $0.367$ & $-0.29$ \\
\hline
-0.424 & $3.0$  & $0.43$ & $0.357$ & $-0.25$ \\
\hline
-0.279 & $2.3$  & $0.36$ & $0.346$ & $-0.20$ \\
\hline
-0.137 & $1.6$ & $0.31$ & $0.336$ & $-0.16$ \\
\hline
0.000 & $1.1$  & $0.25$ & $0.327$ & $-0.12$ \\
\hline
\end{tabular}
\caption{Model B1 with $x=1.35$}
\end{center}
\end{table}


\begin{table}[hbt]
\begin{center}
\begin{tabular}{||l|l|l|l|l||}
\hline
$Log(\frac{R}{R_e})$ & $t_{GW}$ & $[\frac{Fe}{H}]$ & $Mg_2$ & $[\frac{Mg}{Fe}]$  \\
\hline
-2.233 & $/$    & $0.55$ & $0.378$ & $-0.32$ \\ 
\hline
-1.929 & $/$    & $0.55$ & $0.378$ & $-0.32$ \\ 
\hline
-1.385 & $14.3$ & $0.55$ & $0.378$ & $-0.32$ \\ 
\hline
-0.963 & $4.5$  & $0.48$ & $0.366$ & $-0.28$ \\ 
\hline
-0.824 & $3.1$  & $0.44$ & $0.358$ & $-0.25$ \\
\hline
-0.602 & $1.8$  & $0.32$ & $0.338$ & $-0.17$ \\
\hline
-0.424 & $1.0$  & $0.23$ & $0.323$ & $-0.10$ \\
\hline
-0.279 & $0.78$  & $0.17$ & $0.312$ & $-0.060$ \\
\hline
-0.137 & $0.56$ & $0.071$ & $0.296$ & $0.0030$ \\
\hline
0.000 & $0.43$  & $0.025$ & $0.288$ & $0.030$ \\
\hline
\end{tabular}
\caption{Model B2 with $x=1.35$}
\end{center}
\end{table}


\begin{table}[hbt]
\begin{center}
\begin{tabular}{||l|l|l|l|l||}
\hline
$Log(\frac{R}{R_e})$ & $t_{GW}$ & $[\frac{Fe}{H}]$ & $Mg_2$ & $[\frac{Mg}{Fe}]$  \\
\hline
-2.233 & $/$    & $0.55$ & $0.378$ & $-0.32$ \\ 
\hline
-1.929 & $/$    & $0.55$ & $0.378$ & $-0.32$ \\ 
\hline
-1.385 & $/$    & $0.55$ & $0.378$ & $-0.32$ \\ 
\hline
-0.963 & $/$    & $0.55$ & $0.378$ & $-0.32$ \\ 
\hline
-0.824 & $/$    & $0.55$ & $0.378$ & $-0.32$ \\ 
\hline
-0.602 & $/$    & $0.55$ & $0.378$ & $-0.32$ \\ 
\hline
-0.424 & $/$    & $0.55$ & $0.378$ & $-0.32$ \\ 
\hline
-0.279 & $11.9$ & $0.54$ & $0.377$ & $-0.32$ \\
\hline
-0.137 & $7.1$  & $0.52$ & $0.373$ & $-0.31$ \\
\hline
0.000 & $3.9$   & $0.47$ & $0.364$ & $-0.27$ \\
\hline
\end{tabular}
\caption{Model C1 with $x=1.35$}
\end{center}
\end{table}


\begin{table}[hbt]
\begin{center}
\begin{tabular}{||l|l|l|l|l||}
\hline
$Log(\frac{R}{R_e})$ & $t_{GW}$ & $[\frac{Fe}{H}]$ & $Mg_2$ & $[\frac{Mg}{Fe}]$  \\
\hline
-2.233 & $/$    & $0.55$ & $0.378$ & $-0.32$ \\ 
\hline
-1.929 & $/$    & $0.55$ & $0.378$ & $-0.32$ \\ 
\hline
-1.385 & $/$    & $0.55$ & $0.378$ & $-0.32$ \\ 
\hline
-0.963 & $/$    & $0.55$ & $0.378$ & $-0.32$ \\ 
\hline
-0.824 & $/$    & $0.55$ & $0.378$ & $-0.32$ \\ 
\hline
-0.602 & $/$    & $0.55$ & $0.378$ & $-0.32$ \\ 
\hline
-0.424 & $11.3$ & $0.54$ & $0.376$ & $-0.32$ \\ 
\hline
-0.279 & $5.2$ &  $0.50$ & $0.369$ & $-0.29$ \\
\hline
-0.137 & $2.5$ &  $0.40$ & $0.352$ & $-0.22$ \\
\hline
0.000 & $1.1$  &  $0.25$ & $0.327$ & $-0.12$ \\
\hline
\end{tabular}
\caption{Model C2 with $x=1.35$}
\end{center}
\end{table}


\begin{table}[hbt]
\begin{center}
\begin{tabular}{||l|l|l|l|l||}
\hline
$Log(\frac{R}{R_e})$ & $t_{GW}$ & $[\frac{Fe}{H}]$ & $Mg_2$ & $x$  \\
\hline
-2.233 & $/$     & $0.68$ & $0.400$ & $0.95$ \\ 
\hline
-1.929 & $/$     & $0.68$ & $0.400$ & $0.95$ \\ 
\hline
-1.385 & $12.3$  & $0.68$ & $0.400$ & $0.95$ \\ 
\hline
-0.963 & $4.4$   & $0.64$ & $0.393$ & $0.97$ \\ 
\hline
-0.824 & $3.1$   & $0.61$ & $0.389$ & $0.98$ \\
\hline
-0.602 & $1.8$   & $0.54$ & $0.376$ & $1.02$ \\
\hline
-0.424 & $1.2$   & $0.47$ & $0.364$ & $1.05$ \\
\hline
-0.279 & $0.80$  & $0.42$ & $0.350$ & $1.09$ \\
\hline
-0.137 & $0.58$  & $0.39$ & $0.333$ & $1.13$ \\
\hline
0.000  & $0.48$  & $0.17$ & $0.312$ & $1.21$ \\
\hline
\end{tabular}
\caption{Results related to 10-zones IV model.
The IMF is radius dependent in order to have
$[Mg/Fe]$ constant ($=0.15$).} 
\end{center}
\end{table}

\newpage

\section*{Figure Captions}



\noindent
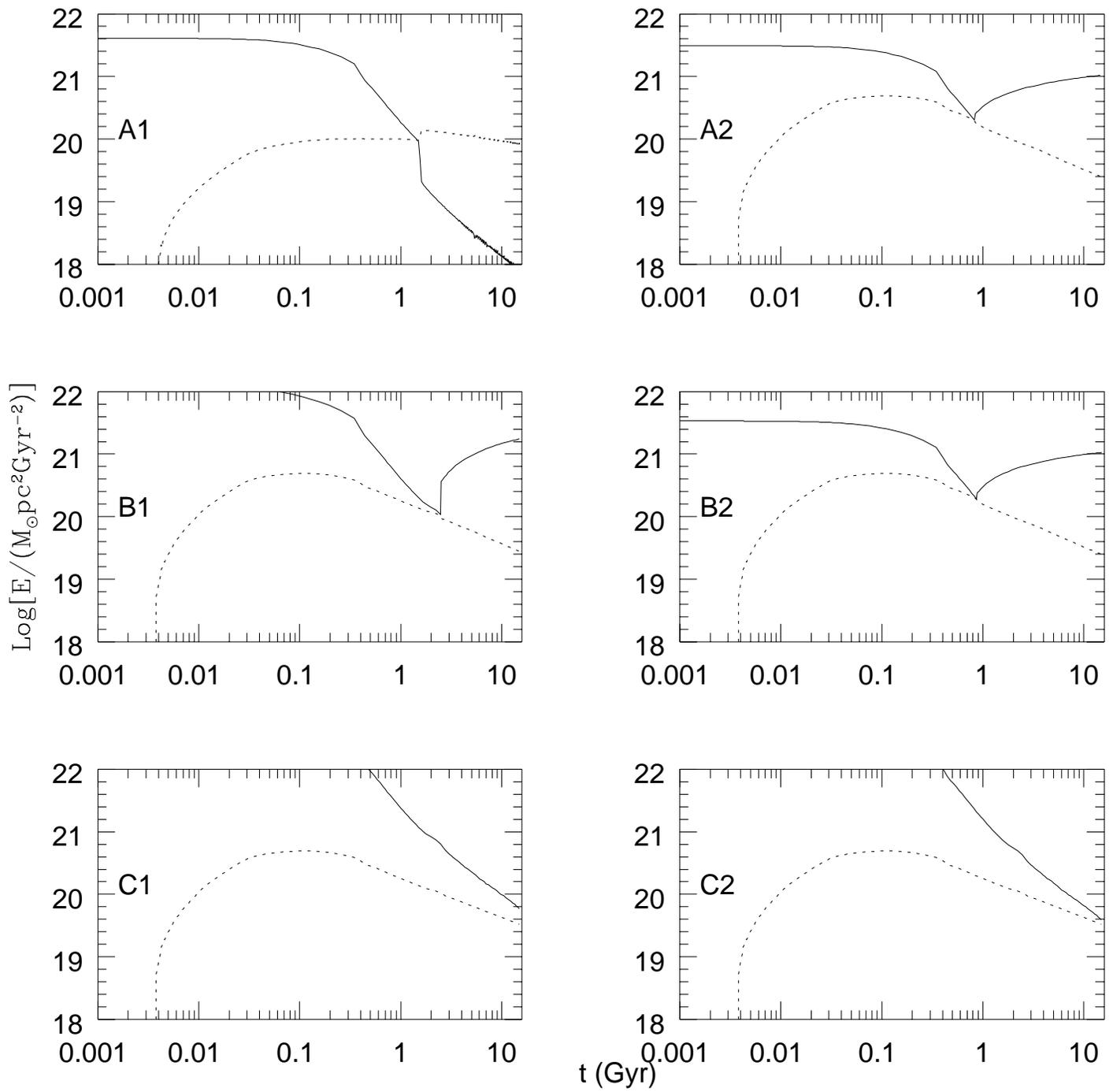
{\bf Figure 1} $E_{Bgas}$ (solid line) and
$E_{th_{SN}}$ (dashed line) for the 6 models in the case of
one-zone.

\noindent
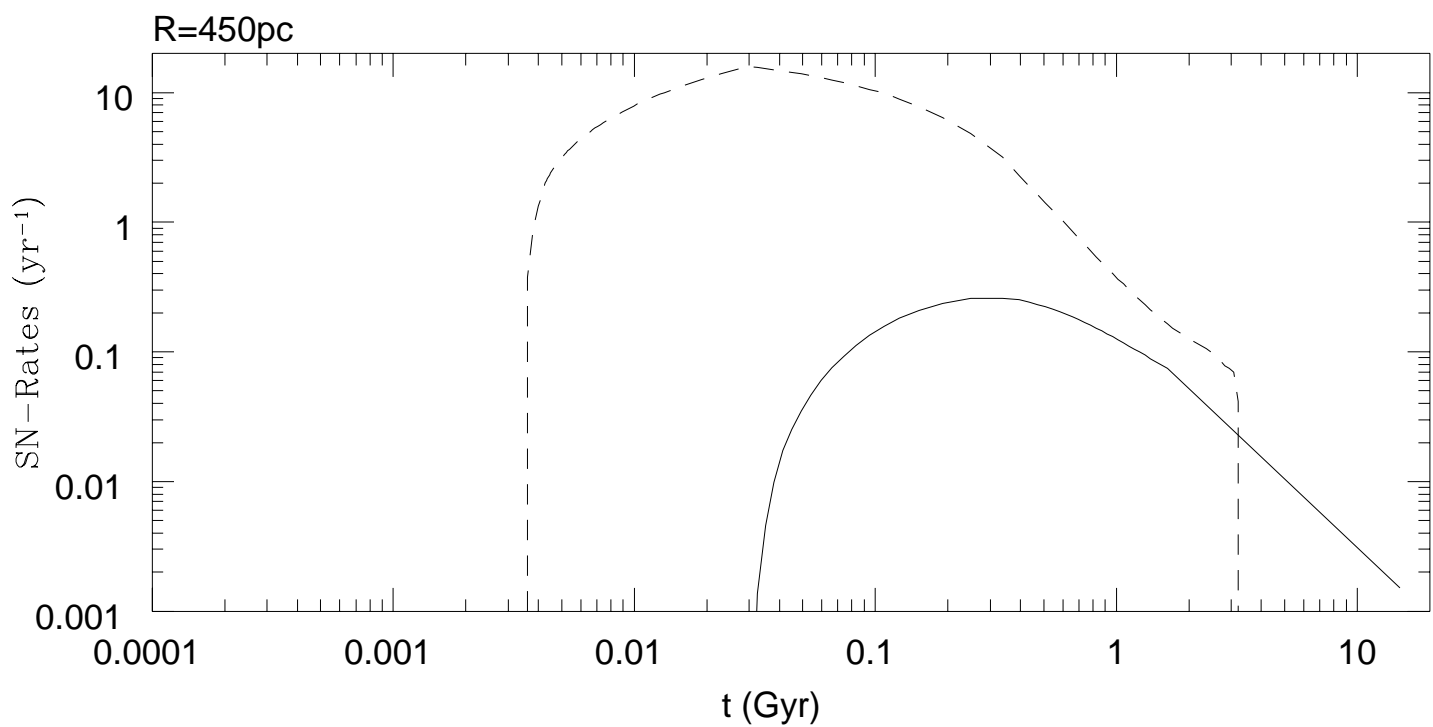
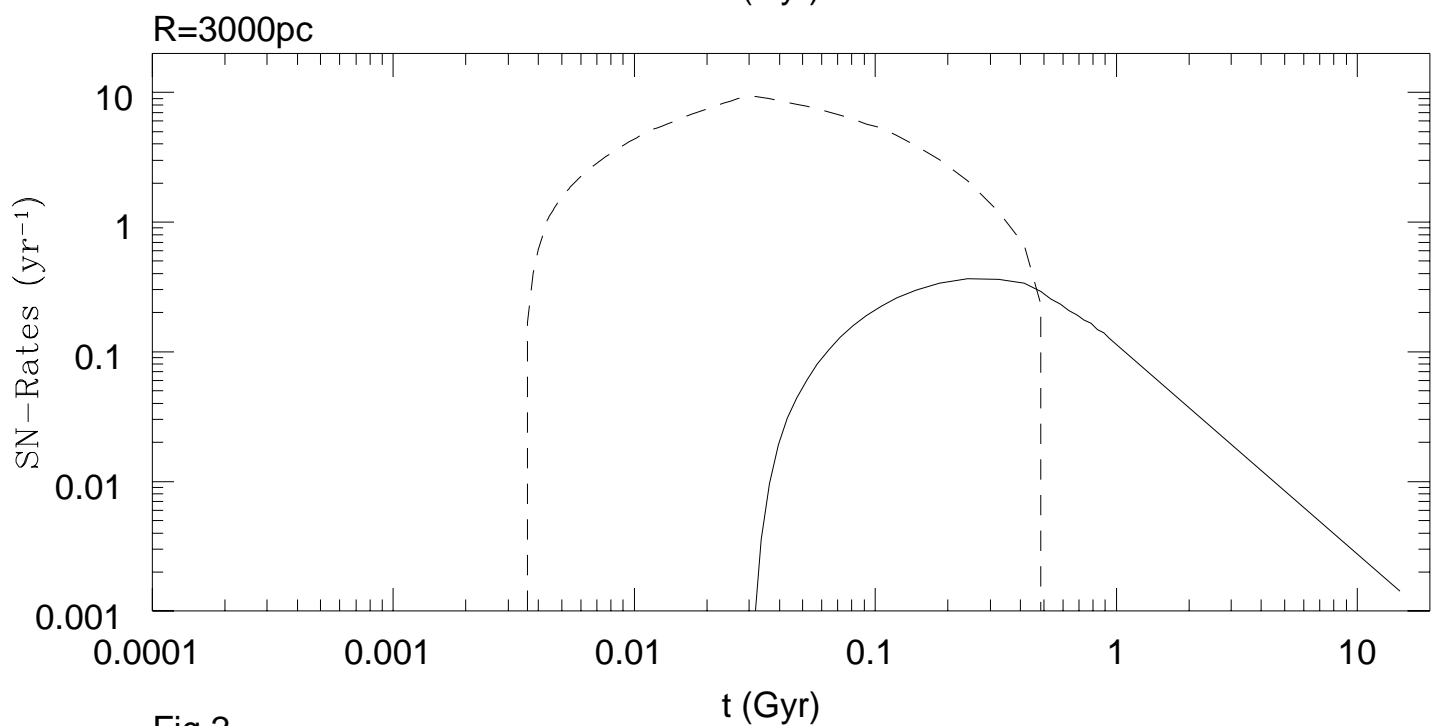
{\bf Figure 2} 
The type II (dashed line) and Ia (continuous line) 
SN rates, expressed in units of SNe $yr^{-1}$,
as a function of time (expressed in Gyr) as predicted by model IV.
The upper panel refers to the central region whereas the lower panel to 
the more external one (corresponding to an effective radius).

\noindent
{\bf Figure 3}
Predicted and observed gradients of the metallicity index $Mg_{2}$.
The data refer to galaxies in the sample observed by Davies et al. (1993).
The continuos line refers to model IV, the dashed line to model B2 (x=1.35)
and the dotted line to model B2 (x=0.95). The error bars in the data are also shown.
The calibration adopted to transform [F/H] into $Mg_2$ is the one of Worthey (1994)
relative to an age of 17 Gyr.

\noindent
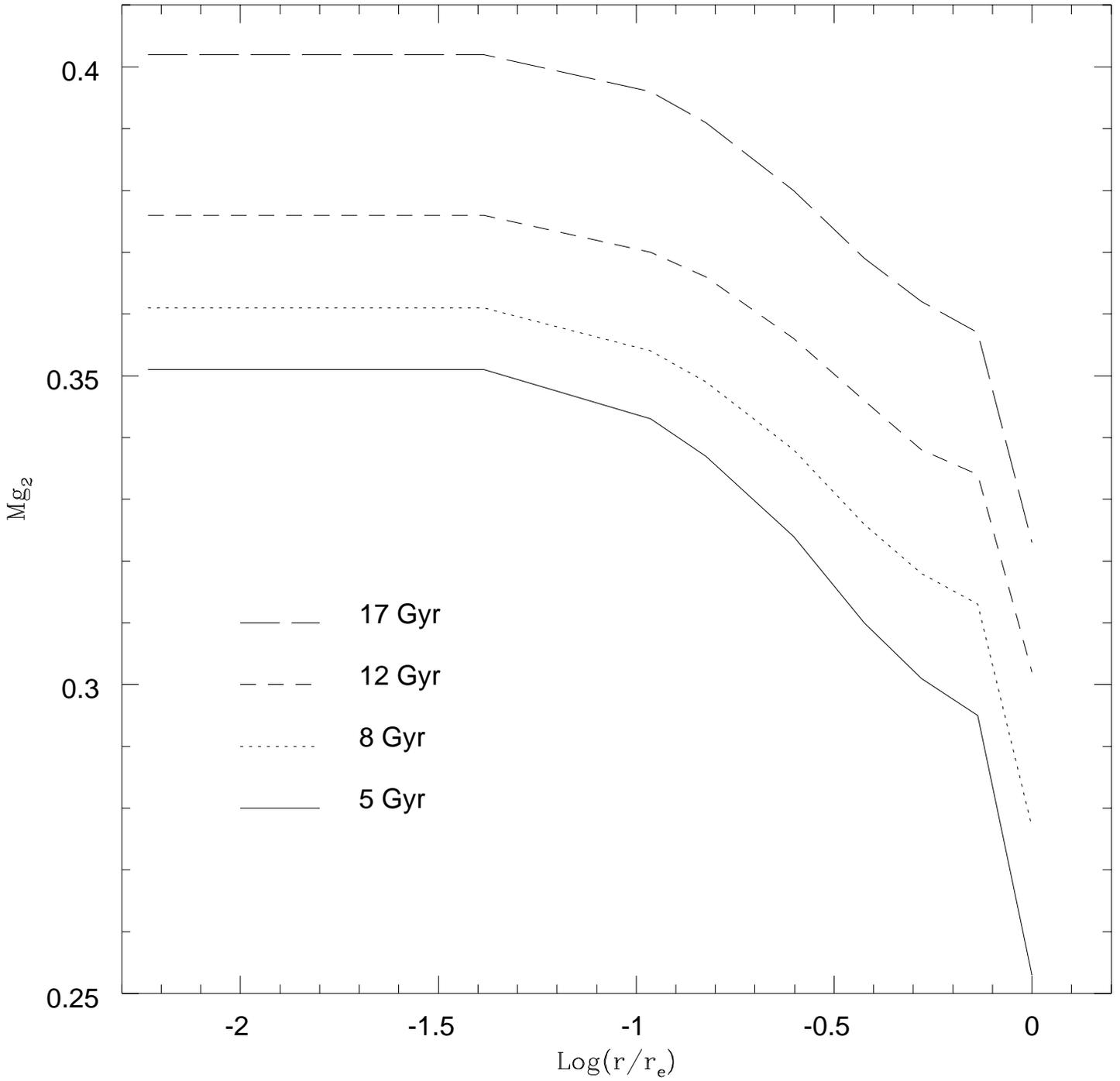
{\bf Figure 4}
Gradients of the index $Mg_2$ as predicted by our model IV by using
the calibrations of Worthey (1994) referring to different ages
of the dominant stellar population.

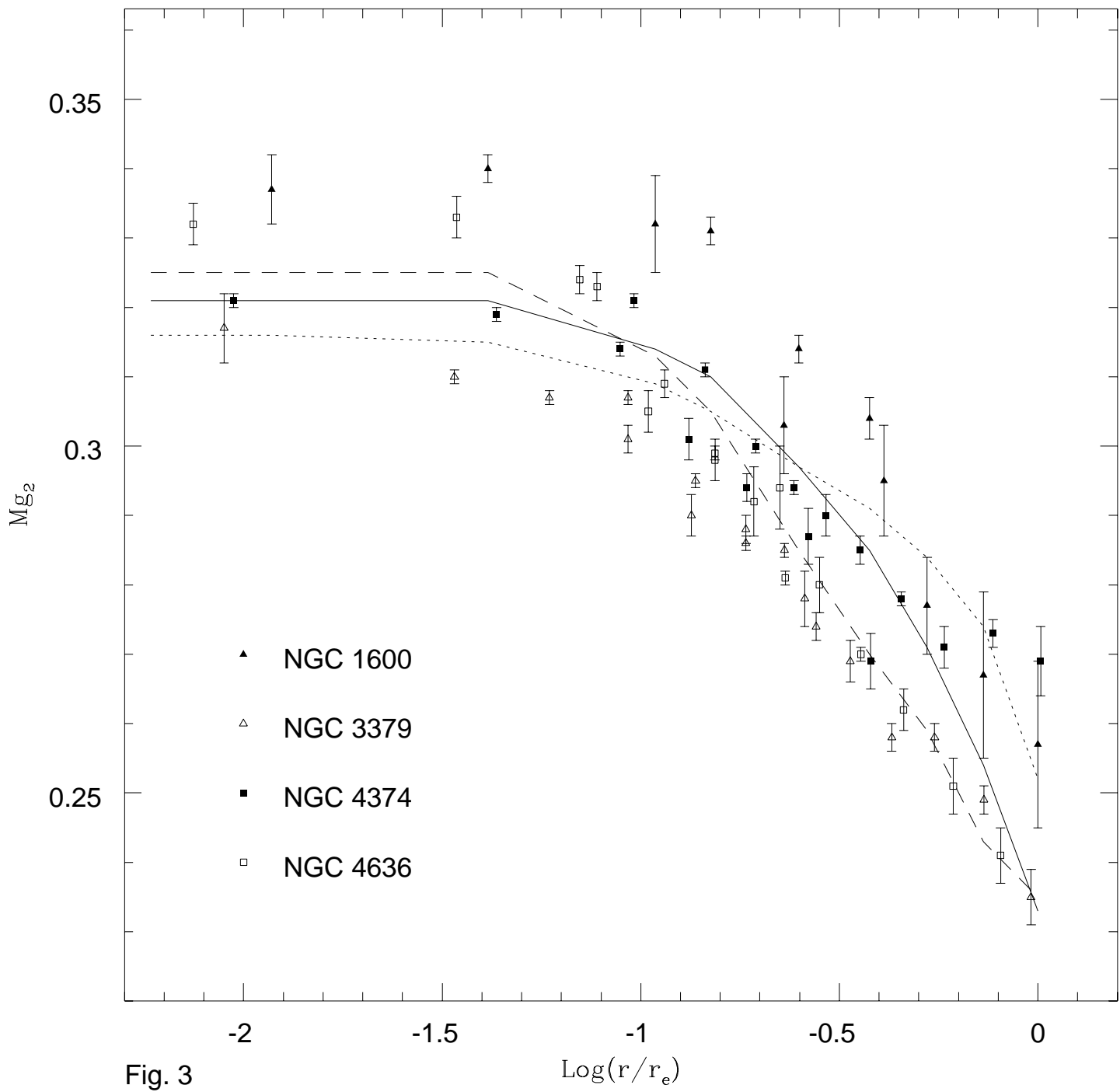
\end{document}